\documentclass[traditabstract]{aa} 
\usepackage{graphicx}
\usepackage{longtable}
\usepackage{rotating}
\usepackage{natbib}
\usepackage[usenames]{color}
\usepackage{multirow}
\usepackage{url}
\usepackage{ulem}
\bibpunct{(}{)}{;}{a}{}{,} 

\newcommand{\ratioo} {N({\rm H}_2) / I_{\rm CO}}

\newcommand{\kms}   {{\rm \  km \  s^{-1}}}
\newcommand{\K}   {{\rm \  K}}

\newcommand{\NHmol} {N(\mathrm{H_{2}})}

\newcommand{\Xunit} {\,{\rm cm^{-2}/(K\kms)}}
\newcommand{\cmt} {\,{\rm cm^{-2}}}
\newcommand{\cc} {\,{\rm cm^{-3}}}
\newcommand{\mum} {{\rm $\mu m$}}
\newcommand{\micron}{\mbox{$\mu$m}}   
\newcommand{\HI}{\ion{H}{i}}   
\newcommand{\OI}{[\ion{O}{i}]}   
\newcommand{\CI}{[\ion{C}{i}]}   
\newcommand{\CII}{[\ion{C}{ii}]}   
\newcommand{\HII}{\ion{H}{ii}}

\usepackage{txfonts}

\begin{document}
  \title{Spectrally resolved CII emission in M~33 ({\tt HerM33es}) }
  \titlerunning{Spectrally resolved CII emission in M~33}
  \subtitle{Physical conditions and kinematics around BCLMP 691 \thanks{{\it Herschel} is an ESA space observatory with science instruments
provided by European-led Principal Investigator consortia and with
important participation from NASA.}}

  \author{J. Braine \inst{1,2} \and P. Gratier \inst{3}  \and C. Kramer\inst{4} \and F. P. Israel\inst{5}  \and  F. van der Tak \inst{6} \and B. Mookerjea\inst{7} \and M. Boquien \inst{8} \and F. Tabatabaei \inst{9} \and P.~van der Werf  \inst{5} \and C.~Henkel \inst{10,11} }

  \institute{Univ. Bordeaux, Laboratoire d'Astrophysique de Bordeaux, F-33270, Floirac, France.\\
             \email{braine@obs.u-bordeaux1.fr}
        \and
  CNRS, LAB, UMR 5804, F-33270, Floirac, France
        \and
           IRAM, 300 Rue de la Piscine, F-38406 St Martin d'H\`eres, France
        \and
            Instituto Radioastronomia Milimetrica (IRAM), 
    Av. Divina Pastora 7, Nucleo Central, E-18012 Granada, Spain
        \and
            Leiden Observatory, Leiden University, PO Box 9513, NL 2300 RA Leiden, The Netherlands 
        \and
            SRON Netherlands Institute for Space Research, Landleven 12, 
    9747 AD Groningen, The Netherlands 
        \and               
	  Tata Institute of Fundamental Research, Homi Bhabha Road,
Mumbai 400005, India 
        \and
           Laboratoire d'Astrophysique de Marseille -- LAM, Universit\'e d'Aix-Marseille \& CNRS UMR~7326, 38 rue F. Joliot-Curie, 13388 Marseille CEDEX 13, France    
        \and
            Max-Planck-Institut f\"ur Astronomie, K\"onigstuhl 17, 69117-Heidelberg, Germany 
        \and
            Max-Planck-Institut f\"ur Radioastronomie (MPIfR), Auf dem H\"ugel 69, D-53121 Bonn, Germany 
        \and Astron. Dept., King Abdulaziz University, P.O. Box 80203,
         Jeddah, Saudi Arabia 
}
  \date{}

 \abstract{ This work presents high spectral resolution observations
   of the \CII\ line at 158~\micron, one of the major cooling lines of
   the interstellar medium, taken with the HIFI heterodyne spectrometer on
   the Herschel satellite.  In BCLMP~691, an \HII\ region far north (3.3 kpc) in
   the disk of M~33, the \CII\ and CO line profiles show similar
   velocities within $0.5 \kms$, while the \HI\ line velocities are
   systematically shifted towards lower rotation velocities by $\sim 5
   \kms$.  Observed at the same $12''$ angular resolution, the \CII\ lines are
   broader than those of CO by about 50\% but narrower than the
   \HI\ lines.  The \CII\ line intensities also follow those of CO
   much better than those of \HI.   A weak shoulder on the \CII\ line
   suggests a marginal detection of the [$^{13}$\ion{C}{ii}] line,
   insufficient to constrain the \CII\ optical depth.  The velocity
   coincidence of the CO and \CII\ lines and the morphology at optical/UV wavelengths indicate that
   the emission is coming from a molecular cloud behind the
   \HII\ region.  The relative strength of \CII\ with respect to the
   FIR continuum emission is comparable to that observed in the
   Magellanic Clouds on similar linear scales but the CO emission 
   relative to \CII\ is stronger in M~33.  The \CII\ line to far-infrared
   continuum ratio suggests a photoelectric heating efficiency of
   1.1\%.  The data, together with
   published models indicate a UV field $G_0 \sim 100$ in units of the
   solar neighborhood value, a gas density $n_H \sim 1000 \cc$, and a
   gas temperature $T\sim 200$ K.  
   Adopting these values, we estimate the C$^+$ column density to be
$N_{C^+} \approx 1.3 \times 10^{17} \cmt$.  The \CII\ emission comes 
predominantly from the warm neutral region between the \HII\ region and 
the cool molecular cloud behind it.  From published abundances, the inferred C$^+$
column corresponds 
to a hydrogen column density of $N_H \sim 2 \times 10^{21} \cmt$.  
The CO observations suggest that $N_H = 2 N_{H_2} \sim 3.2 \times 10^{21} \cmt$ and 21cm measurements, also at $12''$ resolution, yield $N_\HI \approx 1.2 \times 10^{21} \cmt$ within the \CII\ velocity range.  Thus, some H$_2$ not detected in CO must be present, in agreement with earlier findings based on the SPIRE 250 -- 500 $\mu$m emission.
 }
 

\maketitle

\section{Introduction}

The \CII\ line at 157.741~$\mu$m is the strongest emission line from
galaxies, representing close to one per cent of the bolometric luminosity
\citep{Crawford85,Stacey91,Malhotra01} and as such is detectable in
very distant objects \citep{Loeb93,Maiolino05}.  However, the question of the
origin of the \CII\ emission from galaxies \citep{Heiles94} remains open
to this day.  Possible sources are the diffuse warm ionized
medium, \HII\ regions, and mostly neutral photo-dissociation regions
(PDRs).  Because carbon is ionized more easily than hydrogen (11.26eV
rather than 13.6eV), \CII\ emission can originate in otherwise neutral
gas.  A series of \CII\ observations in the Local Group galaxy M~33
with the HIFI (Heterodyne Instrument for Far Infrared astronomy)
instrument on the Herschel Space Observatory
\citep{Pilbratt10,deGraauw10,Roelfsema12} is the essential component
of the Open Time Key Project HerM33ES \citep{Kramer10}.  A first
article \citep{Mookerjea11} presented a PACS map of the \CII\ and
\OI\ emission and a HIFI \CII\ spectrum of the \HII\ region BCLMP 302.
This work is the first of a series presenting extended high spectral resolution
observations of the \CII\ line in M~33.

Despite its importance, only a few velocity-resolved observations have
been made of the \CII\ line \citep[e.g. ][]{Boreiko91}.  The goal of such
observations is not only to obtain line intensities but also velocity
emission profiles to enable a comparison of the velocities and
velocity widths of the \CII-emitting gas with those of the lines due
to CO and HI (and H$\alpha$ in the future).  The \CII\ line emission
provides information on the heating and cooling processes of the gas
and is a rather unique probe of the photon-dominated region (PDR)
between the cool CO-emitting molecular gas and the gas ionized by
massive stars.  This region is likely to be more massive and extended
in sub solar-metallicity galaxies and may represent a significant but
hitherto unmeasured gas mass.  Hyper-fine components of the
[$^{13}$\ion{C}{ii}] line bracket the main isotope \citep{Cooksy86}
and with HIFI we are, in principle, able to separate the lines
allowing a direct determination of the optical depth.  

\begin{figure}[!h!]
\begin{flushleft}
\includegraphics[angle=0,width=8.5cm]{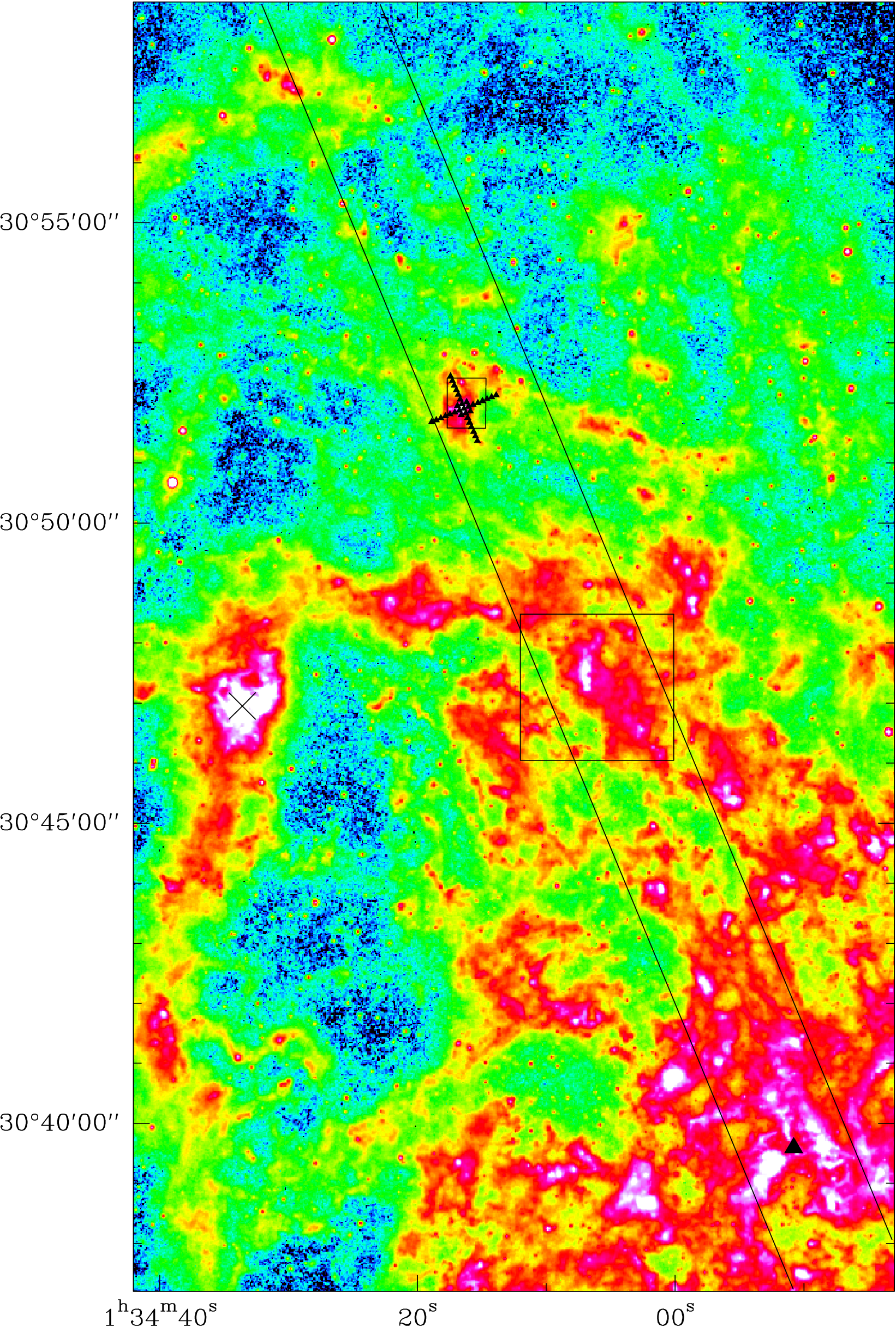}
\caption{\label{b691_fig1} North-eastern part of M33 in the IRAC
  8~\micron\ band \citep{Verley07}.  The upper (small) rectangle shows the
  region presented in Fig.~\ref{superpose_spec} and the small
  triangles forming an 'X' in and around the rectangle indicate the
  individual positions of the \CII\ HIFI observations.  The large
  lower rectangle shows the region around BCLMP302 observed with PACS
  by \citet{Mookerjea11}.  The center of M~33 is at the position of
  the triangle to the lower right of the image of M~33.  The giant
  \HII\ region NGC~604 can be seen as the very bright region to the
  left around Dec 30:47 and is marked with a $\times$.}
\end{flushleft}
\end{figure}

Recently, \citet{Langer10}  found a population of diffuse
interstellar clouds ($A_V < 1.3$ mag) in the Milky Way which show
\CII\ and \HI\ emission, but no detectable CO emission.  The observed
\CII\ emission is stronger than expected for diffuse atomic clouds.
The idea of CO-dark H$_2$ is not new and  
(a) some is definitely found around CO-bright clouds \citep[e.g.][]{Pak98,Wolfire10} 
but possibly also (b) mixed with the \HI\ \citep[e.g.][]{Papadopoulos02} or 
(c) in the outermost parts of galactic disks \citep[e.g.][]{Pfenniger94a}.
\citet{Langer10} attribute the \CII\ excess emission to diffuse and
warm (T$\ga 100$ K) molecular (H$_2$) clouds. 
While the linear resolution of these
observations is excellent, the Galactic plane is optically thick at
short wavelengths such as the H$\alpha$ line commonly used to trace \HII\ regions.  
Thus, some of the \CII\ emission
they detect may in fact be due to ionized rather than neutral gas.
The observations presented here allow us to address the same questions
with $\sim~ 50$pc resolution but with the advantage of much clearer
views of the region and of the sources of emission at all wavelengths.

Previous observations have established that \CII\ emission is a good
tracer of star formation regions.  Among Local Group galaxies, there is an
excellent correspondence between \CII\ and Far-IR emission in both
Magellanic Clouds \citep{Poglitsch95,Israel96,Israel11} 
and between \CII\ and both H$\alpha$ and mid-IR (24$~\mu$m )
emission in the spiral arms of M~31 \citep{Rodriguez06}. The fact that the
\CII\ line represents a similar fraction of the global FIR luminosity of
spiral galaxies shows that this also holds in a statistical way on galactic
scales.

As illustrated in Fig. 1, the HerM33es project observes \CII\ in a long strip along the major
axis. In this work, we investigate the relationship between \CII\,
FIR, CO, and \HI\ emission at the scale of individual clouds ($\sim$50
pc) in an area centered on the \HII\ region BCLMP~691
\citep[]{Boulesteix74} and compare it with the results from
\citet{Mookerjea11} on BCLMP~302, both part of the strip.  In
particular, we estimate the amount of gas traced by the \CII\ emission
and and relate that to the CO and HI emission. 

\begin{figure*}[!h!]
\begin{flushleft}
\includegraphics[angle=0,width=18cm]{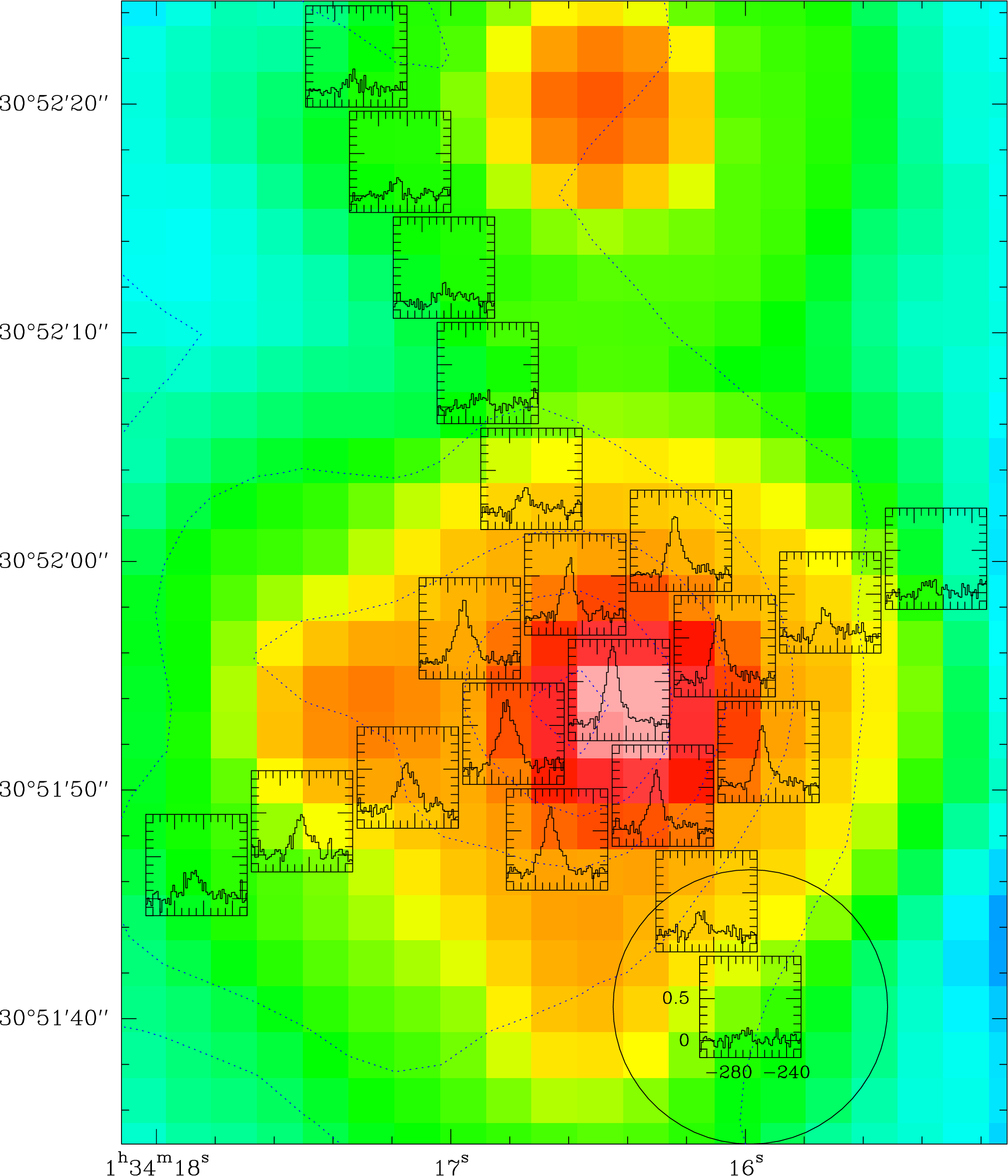}
\caption{\label{superpose_spec} \CII\ spectra superposed on a Spitzer
  24~\micron\ image at $6"$ resolution of BCLMP~691.  The dotted blue contours show the
  H$\alpha$ emission at levels of 100, 300, 1000, 3000, and 10000
  counts \citep{Greenawalt98,Hoopes00}.  Only positions with detected
  \CII\ flux are shown. The velocity scale is from $-300$ to $-230
  \kms$ and the $y$ axis is from -0.2~K to 1~K.  The beamsize is shown
  as a circle around the southernmost spectrum.}
\end{flushleft}
\end{figure*}

\section{Observations and data reduction}

The HIFI Wide Band Spectrometer horizontal and vertical polarization
data were processed to level 2 using the standard pipeline of HIPE
version 4 and then exported to CLASS (see http://www.iram.fr/IRAMFR/GILDAS).
The obsids for the BCLMP691 observations are 1342213155 and 1342213156
which are load-chop On-The-Fly maps consisting of 8280 individual
spectra each.  The maps are perpendicular strips with position angles
of 22.5 and 112.5 degrees with close spacing between individual
spectra.  All spectra were smoothed to 8 MHz ($1.26 \kms$) resolution
and a zero-order baseline (i.e. a continuum level) was subtracted
excluding velocities from $-280$ -- $-255 \kms$, corresponding to the
expected line velocity.  The baseline was calculated over a region
extending from $-345$ to $-170 \kms$.  All velocities are in the LSR 
frame and the systemic velocity of M~33 is about $-170\kms$.

A number of spectra (roughly 1/3) suffer from the HIFI HEB (hot
electron bolometer) standing wave pattern at about 290 MHz.  Various
Fourier filtering schemes were applied but it proved very difficult to
reliably eliminate the standing waves.  In the end, the 11298 good
spectra were kept and the others dropped.

At 5$"$ intervals along the cross, the spectra were convolved with a
5$"$ gaussian in order to obtain a single high quality spectrum at
each position. The resulting beam size is 12.3$\times 11.4"$ because
the convolution only stretches the beam in the direction where there
are spectra (i.e. along the lines of the cross).  In the center, the
beam is more circular because both arms of the cross are included in
the convolution.  Four "extra" spectra were generated near the center
at ($\pm 5",\pm 5"$) with respect to the cross (see
Fig.~\ref{superpose_spec}).  Although no pointings correspond exactly
to these positions, this is possible because the neighboring beams
overlap considerably in this region so these points are sampled on
both arms of the cross.

\begin{table*}[h*]   
\caption[]{ Positions observed in \CII.  Columns 1 and 2 give the
  offset from the nominal pointing center of (J2000) RA: 01:34:16.40
  DEC: 30:51:54.6 in arcseconds (B1950 coordinates are RA: 01:31:26.98
  DEC: 30:36:33.6).  Columns 3, 4, and 5 give \CII\ fluxes, velocities
  and line widths for the detected positions.  For the non-detections,
  column 3 gives the $rms$ noise level in mK.  CO integrated
  intensities, velocities, and line widths are provided in columns 6
  -- 8 and the \HI\ column density is indicated in column 9.
  Velocities and line widths (and their uncertainties) are estimated
  via gaussian fits to the spectra; when no value is present, the
  routine did not converge or the spectrum was not considered a
  detection.  The CO integrated intensities and \HI\ column densities
  are taken from \citet{Gratier10}.  Finally, the 24~\micron\ flux
  density \citep{Tabatabaei07a} is given in the last column.}
\begin{tabular}{lrrrrrrrrrr}   
\hline \hline  
$x(")$&$y(")$& $I_{\CII}$ & $v_{\CII}$ & $\Delta v_{\CII}$ & $I_{CO}$ & $v_{CO}$ & 
$\Delta v_{CO}$ & N$_{\HI}$ & $S_{24}$   \\
&&K km s$^{-1}$&km s$^{-1}$&km s$^{-1}$&K km s$^{-1}$&km s$^{-1}$&km s$^{-1}$&$10^{21}$ cm$^{-2}$ & MJy sr$^{-1}$ \\
\hline
18.5& -7.7&  4.5$\pm$0.4& -266.7$\pm$ 0.7&
14.8$\pm$1.8&  1.8& -265.9$\pm$0.4&   8.0$\pm$1.0&  1.89&   2.4\\
13.9& -5.7&  5.9$\pm$0.4& -265.1$\pm$ 0.4&
13.4$\pm$1.0&  3.0& -266.1$\pm$0.2&   7.6$\pm$0.6&  1.98&   6.1\\
-5.7&-13.9&  1.1$\pm$0.3& -268.6$\pm$ 1.1&
 8.8$\pm$2.3&  1.7& -268.7$\pm$0.3&   7.1$\pm$0.7&  1.41&   1.4\\
 9.2& -3.8&  8.2$\pm$0.4& -265.1$\pm$ 0.4&
15.3$\pm$1.0&  4.0& -266.9$\pm$0.1&   7.8$\pm$0.4&  1.90&  11.0\\
-3.8& -9.2&  2.5$\pm$0.3& -268.9$\pm$ 0.6&
 9.4$\pm$1.4&  2.6& -269.0$\pm$0.3&   7.3$\pm$0.7&  1.43&   5.1\\
 4.6& -1.9& 11.0$\pm$0.4& -269.0$\pm$ 0.3&
14.5$\pm$0.6&  4.3& -269.0$\pm$0.2&   8.1$\pm$0.4&  1.84&  45.5\\
-1.9& -4.6&  6.7$\pm$0.3& -269.6$\pm$ 0.2&
10.0$\pm$0.6&  2.6& -269.5$\pm$0.2&   7.1$\pm$0.5&  1.39&   8.7\\
 0.0&  0.0&  9.9$\pm$0.2& -269.4$\pm$ 0.1&
10.7$\pm$0.3&  4.0& -269.9$\pm$0.2&   7.1$\pm$0.4&  1.66&  46.3\\
-4.6&  1.9&  7.8$\pm$0.3& -269.1$\pm$ 0.2&
10.6$\pm$0.5&  2.8& -269.4$\pm$0.2&   7.2$\pm$0.6&  1.39&   8.4\\
 1.9&  4.6&  6.9$\pm$0.3& -269.1$\pm$ 0.2&
10.4$\pm$0.6&  3.7& -269.3$\pm$0.2&   6.9$\pm$0.4&  1.68&  19.8\\
-9.2&  3.8&  3.2$\pm$0.3& -268.3$\pm$ 0.5&
10.2$\pm$1.3&  1.9& -268.2$\pm$0.7&  11.9$\pm$2.4&  1.11&   2.7\\
 3.8&  9.2&  3.0$\pm$0.4& -269.2$\pm$ 0.7&
10.8$\pm$1.8&  2.5& -269.1$\pm$0.5&   7.5$\pm$1.5&  1.54&   7.8\\
-13.9&  5.7&  1.2$\pm$0.3& -269.7$\pm$ 1.1&
 8.9$\pm$2.1&  1.3&  &   &  0.82&   0.6\\
 5.7& 13.9&  1.7$\pm$0.3& -269.4$\pm$ 1.4&
12.3$\pm$2.7&  2.2& -266.0$\pm$1.0&  16.8$\pm$2.5&  1.59&   3.3\\
 7.7& 18.5&  1.8$\pm$0.5& -262.5$\pm$ 2.6&
13.6$\pm$5.5&  1.9& -267.9$\pm$0.5&  10.3$\pm$1.3&  1.91&   2.3\\
 9.6& 23.1&  1.5$\pm$0.3& -267.5$\pm$ 0.7&
 7.9$\pm$1.7&  1.9& -267.2$\pm$0.4&   8.3$\pm$1.0&  2.11&   3.7\\
11.5& 27.7&  1.6$\pm$0.4& -267.5$\pm$ 1.0&
 9.7$\pm$3.1&  2.5& -267.6$\pm$0.6&  11.4$\pm$2.2&  2.05&   3.0\\
 2.7& -6.5&  9.2$\pm$0.3& -269.4$\pm$ 0.2&
12.9$\pm$0.5&  3.8& -269.4$\pm$0.2&   7.2$\pm$0.4&  1.72&  10.5\\
-6.5& -2.7&  7.1$\pm$0.3& -269.3$\pm$ 0.2&
10.5$\pm$0.5&  1.5& -269.3$\pm$0.4&   6.5$\pm$0.8&  1.13&   4.7\\
-2.7&  6.5&  7.5$\pm$0.3& -268.8$\pm$ 0.2&
11.1$\pm$0.6&  3.4& -268.6$\pm$0.2&   7.6$\pm$0.5&  1.56&   6.8\\
 6.5&  2.7&  9.1$\pm$0.3& -268.9$\pm$ 0.2&
13.4$\pm$0.5&  4.2& -268.9$\pm$0.2&   8.2$\pm$0.5&  1.75&  27.2\\
32.3&-13.4&  109.7&&
&  0.1&      &     &  1.49&   0.5\\
-13.4&-32.3&   92.4&&
&  0.8&      &     &  1.64&   0.6\\
27.7&-11.5&  125.5&&
&  0.3&      &     &  1.53&   0.9\\
-11.5&-27.7&   67.8&&
&  1.0&      &     &  1.53&   0.6\\
23.1& -9.6&   75.0&&
&  1.0&      &     &  1.59&   1.6\\
-9.6&-23.1&   75.0&&
&  1.4&      &     &  1.42&   0.8\\
-7.7&-18.5&   66.4&&
&  1.5&  -269.4$\pm$0.5 & 7.6$\pm$1.0 &  1.56&   0.8\\
-18.5&  7.7&   60.6&&
&  0.8&      &     &  0.63&   0.4\\
-23.1&  9.6&   57.7&&
&  0.2&      &     &  0.53&   0.4\\
-27.7& 11.5&   77.9&&
&  0.4&      &     &  1.09&   0.5\\
-32.3& 13.4&   82.3&&
&  0.6&      &     &  1.37&   0.7\\
13.4& 32.3&   66.4&&
&  2.1&   -267.8$\pm$0.6  & 10.2$\pm$1.4   &  2.17&   2.0\\
\hline
\end{tabular}   
\end{table*}   

 Velocities, line widths, and line intensities were measured from
 gaussian fits to the 17 detected positions on the cross plus the 4
 "extra" positions.  Table 1 gives the positions of the spectra with
 the fit results for the detected positions and the $rms$ noise level
 for the others.  The uncertainties come from the gaussian fits using
 the CLASS fitting routine.  As per the current HIFI users manual {\tt
   http://herschel.esac.esa.int/Docs/HIFI/html/
   ch05s05.html\#beam-information}, we adopt a main beam efficiency of
 0.693 and present all data on the main beam temperature scale.

\section{Analysis of BCLMP~691 \CII\ line intensities}

\subsection{The \HII\ region complex BCLMP~691}

Both BCLMP~691 (IK~60, Aller~3, Searle~15) and BCLMP~302 (als known as
IK~53, Aller~43, Searle~11) are among the brighter nebulae in the
northern half of M~33 with 1.4 GHz flux densities of 4.6 and 6.1 mJy
respectively (\citep{Boulesteix74}, \citep{Israel74}).  Figure 1
shows the northern part of M~33 with the
region observed around BCLMP~691 (and BCLMP~302) indicated.  Zooming
in on BCLMP~691, Figure \ref{superpose_spec} shows the
\CII\ detections superposed on a 24~\mum\ image \citep{Tabatabaei07a}
of the region.  BCLMP~691 and BCLMP~302 are  similar in many
aspects.  Both are about 75 pc in diameter with large-scale r.m.s electron 
densities of $<n_{e}^2>^{1/2}$ of 7.4 and 6.3 cm$^{-3}$ and excitation 
parameters of $u = R_{Stromgren} n_e^{2/3} = 180$ and 200
pc cm$^{-2}$, corresponding to excitation by the equivalent of 3 and
4 O5 exciting stars \citep{Israel74}. More significant differences
between BCLMP~691, the HII region studied in this paper, and BCLMP~302
(\citep{Mookerjea11}) are tied to their location.  
BCLMP~691 is a major \HII\ region 3.3 kpc from the center, and
relatively isolated, whereas BCLMP~302 is a major star-forming region
embedded in the northern spiral arm at 2.1 kpc from the center of M~33.  
The \HI\ and CO($J$=2-1) emission towards the latter is stronger than 
towards BCLMP~691.  
BCLMP~691 has a metallicity 12+log(O/H) $= 8.42\pm0.06$ 
\citep{Magrini10}, {\it i.e.} an oxygen abundance relative to H of 
$2.6\,\times\,10^{-4}$, about half the metallicity of the solar
neighborhood.  This is somewhat below the average of the more central
\HII\ regions such as NGC~595 \citep{Magrini10}.  BCLMP~302 is
irradiated by the combined young and old stellar population of the
northern spiral arm, but in the vicinity of BCLMP~691, the radiation
field from the old stellar population is weak. Nevertheless, the
difference in ionization between the two regions is remarkable: the
isolated complex BCLMP~691 has a much higher degree of ionization than
BCLMP~302, with ratios Ne$^{\rm ++}$/Ne$^{\rm +}$ = 1.25 (0.19) and
S$^{\rm 3+}$/S$^{\rm ++}$ = 0.15 (0.04) where the values in
parentheses are those of BCLMP~302 \citep{Rubin08}. The ionization
difference may be attributable to the lower metallicities at greater radii
causing the exciting stars to have harder radiation fields.  
As can be seen in the appendix of \citet{Gratier11}, the geometry of the 
\HII\ region -- molecular cloud interface differs between BCLMP~302 and 691
\citep[clouds 256 and 299 of ][]{Gratier11}.  It is
clearly of interest to identify the corresponding effects on the \CII\ emission.

\subsection{The \CII\ to FIR and the \CII\ to CO ratios}

The large-scale \CII\ to FIR continuum flux ratio in normal galaxies
is usually below 1\%, typically 0.3 -- 0.5\% \citep{Crawford85,
  Stacey91,Braine_n4414c,Malhotra01}.  Here, FIR refers to the dust
emission in the wavelength range 42-122$\mu$m (IRAS definition). This
is about half of all dust emission \citep[i.e. FIR $\approx$ 1/2 TIR, see][]{Dale02}.
At the scale of a spiral arm in M~31, using ISO LWS data at 300pc
resolution, \citet{Rodriguez06} find a considerably higher ratio 
(I$_{\CII}$/FIR$_{42-122} \approx 2$\%).  The closest to M~33 in metallicity is the
LMC, where \citet{Israel96} and \citet{Israel11} find I$_{\rm CII}$/FIR flux 
ratios between of 0.7 and 5.0~\%. 
The \CII\ flux can be calculated as 
\begin{equation}
 I_{\CII} = \frac{2 k \nu^3} {c^3} \, 10^5 \, \int{Tdv} = 7.035 \times 10^{-6} \, \int{Tdv}
\end{equation} 
where $\int{Tdv}$ is the integrated intensity in the standard units of K$\kms$
(the $10^5$ converts from K$\kms$ to K cm s$^{-1}$).

 For BCLMP~691 we
find a ratio of I$_{\CII}$/FIR$_{42-122} \approx 1.1$\% and a ratio of \CII\ flux to 
total dust emission of 0.54\%.  The I$_{\CII}$/IR ratio is indicative of the gas heating efficiency
\citep{Bakes94} and the value for BCLMP~691 is very close to the average ratio
obtained by \citet{Mookerjea11} for BCLMP~302 (their region A).  
In this discussion we focus on the center of the \HII\ region, our (0,0) position.
The FIR and total dust
luminosities for this region were calculated by directly summing the emission
in these wavebands using the 5.8, 24, and 70 $\mu$m emission from Spitzer and 
the PACS and SPIRE data obtained by the HerM33es project.  
PAH emission has been left out of the total dust flux. In both M~33 fields
observed, these ratios are thus very similar to each other and to those
obtained in the LMC and the SMC \citep{Israel11}.
 
We may also compare the \CII\ to CO flux ratios for the various
environments.  Unfortunately, we do not have $J$=1-0 $^{12}$CO
measurements at 12$"$ resolution so we have convolved our CO(2--1) 
observations to the 25$"$ resolution of the CO(1--0).  At this 100pc 
resolution, the CO(2--1)/(1--0) line ratio is 0.72, very close to the average
value given in \citet{Gratier10}.  

For BCLMP~691, a comparison of \CII\ and CO lines can be found in
Tab. 1 and the spectra are superposed in Fig.~\ref{spectra}.  
As for \CII, the CO flux can be calculated as
$ I_{CO} = \frac{2 k \nu^3}{c^3} \, 10^5 \, \int{Tdv} = 1.57 \times 10^{-9} \int{T_{CO(1-0)}dv} $ 
where $\int{Tdv} $ is in K$\kms$ as usual.  Assuming 
the CO(2--1)/(1--0) line ratio is also 0.72 at 12$"$ resolution, we find
a I$_{\CII}$/I$_{CO(1--0)}$ flux ratio of about 8000 for BCLMP~691, slightly lower than
the BCLMP~302 peak value (Mookerjea et al).  Although this is in excess
of the ratios of about 1500 found in normal spiral galaxies, and even
of ratios of 6000 such as found in starburst galaxies
\citep{Stacey91,Braine_n4414c}, it is considerably lower than the
corresponding ratios found in the LMC, which (with one exception,
N~159W) range from 13000 to more than 100000 with a mean of 60000 for
the individual star-forming clouds observed, and a ratio of 23000 for
the LMC as a whole \citep{Israel96,Israel11}.

In general, because of its (indirect) dependence on UV photons,
\CII\ emission is linked to star formation.  GMCs without
\HII\ regions have much weaker \CII\ emission for comparable levels of
CO emission, such as N~159S in the LMC with a I$_{\CII}$/I$_{CO}$ ratio of 400
\citep{Israel96}.  In comparison with the LMC clouds, the two objects
in M~33 have rather low I$_{\CII}$/I$_{CO}$ ratios although I$_{\CII}$/FIR ratios are
practically {\it identical}.  Since the latter are a measure of the
heating efficiency of the clouds, the conclusion must be that
otherwise comparable star-forming complexes are much richer in CO in
M~33 than in the LMC.  Since there is no difference in this respect
between the two M~33 objects, the degree of ionization, i.e. the
hardness of the radiation field, is not a factor of significance.

\subsection{Carbon column density}

\subsubsection{A lower limit to N$_{C^+}$ }

The \CII\ line is an important probe of the physical conditions of the
ISM.  In particular, it provides information on regions not probed by
the CO line and possibly not probed by the \HI\ line.  The \CII\ line
is thermalized at a density of about $3000\cc$ in neutral gas and has
an upper level energy of 91~K.  Thus, a minimum \CII\ column
  density can be calculated assuming that densities $\gg 3000\cc$ and
temperatures $\gg 91$K.  Lower densities and/or lower temperatures
would decrease the \CII\ emission per H-atom.  A lower limit to the
hydrogen column density can be determined by assuming a carbon
abundance.  \citet{Israel96} and \citet{Israel11} performed comparable
studies at very similar linear resolutions in the brightest
\HII\ region complexes of the Large Magellanic Cloud, which has a
metallicity similar to that of M~33.  These studies therefore provide
an excellent basis for comparison, even though they did not measure
the actual \CII\ line profiles.  Following the reasoning and equations
presented by \citet{Crawford85} and \citet{Israel96}, 
\begin{equation}
N_{C^+} \ge 6.34 \times 10^{20} \, I_{\CII} \cmt
\end{equation}
with $ I_{\CII}$ defined in Eq. (1).

The \CII\ line intensity at the peak of 9.9~K$\kms$ is equivalent to
$7.0 \times 10^{-5}$ erg $\cmt$s$^{-1}$sr$^{-1}$, or a column density
of $N_{C^+} \ge 4.4 \times 10^{16} \cmt$.  We emphasize once more that
this is a {\it lower} limit because of the high-density,
high-temperature assumption underlying these numbers.  We have also
assumed that all of the ionized carbon was singly ionized -- this
latter hypothesis is reasonable for the neutral medium (i.e. where H
is not ionized).  

That carbon is only singly ionized is not a good approximation 
for regions where \HII\ is the dominant
phase.  The ionization potential for doubly-ionized carbon is 24.38
eV, considerably below that of doubly-ionized neon (40.96 eV) or
triply ionized sulphur (34.83 eV). Given the high degree of ionization
characterizing BCLMP~691 (Section 3.1), in the \HII\ region itself
most carbon will be in the higher ionization states.  Therefore, in
the following, we will compare the derived carbon column density with
that of the neutral gas.  By the same token, we expect little of
the \CII\ emission to be contributed by the ionized gas in the
\HII\ region, which would have caused us to overestimate the amount of
emission from the neutral gas. Rather, we may assume that essentially
all of the \CII\ is indeed associated with the \HI\ and H$_2$.  
As \citet{Mookerjea11} have estimated that emission from
the \HII\ region BCLMP~302 might account for 20 -- 30\% of the
\CII\ emission, this is another potential distinction between that
complex and the one studied here.

\begin{figure}[!h]
\begin{flushleft}
\includegraphics[angle=0,width=8cm]{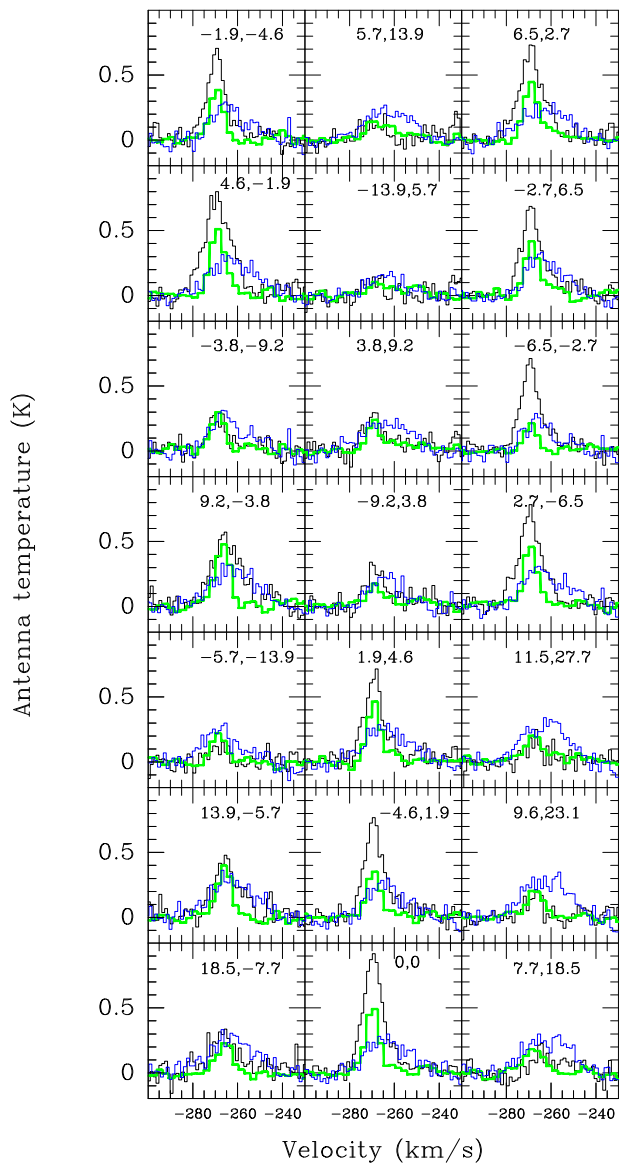}
\caption{\label{spectra} \CII, CO, and \HI\ spectra of all detected
  positions at $12"$ resolution.  Offsets in arcsec are indicated in
  the upper right corner of each panel.  The order follows that of
  Table 1 from bottom left to upper right.  \CII\ is in black, CO in
  green, and \HI\ is in blue.  Y-scale is main beam antenna
  temperature in Kelvins for \CII\ and CO; \HI\ temperatures need to
  be multiplied by 150 to obtain true antenna temperatures.}
\end{flushleft}
\end{figure}
\begin{figure}[!h]
\begin{flushleft}
\includegraphics[angle=0,width=8.3cm]{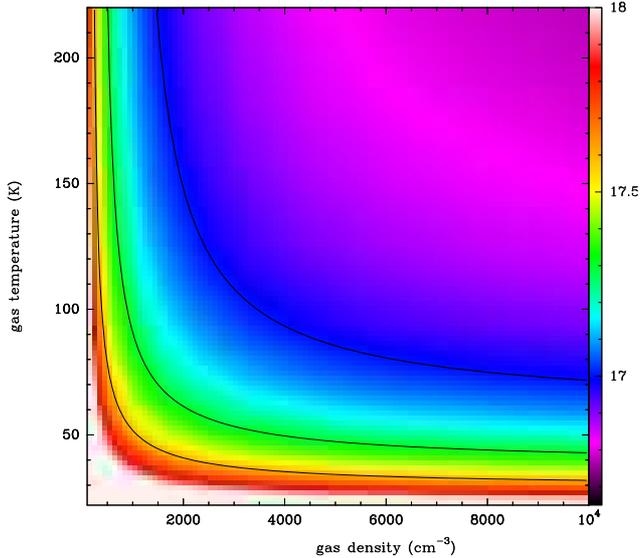}
\caption{\label{ncii} Plot of the log of the C$^+$ column density as a
  function of gas density and temperature for the \CII\ intensity at
  the (0,0) position of $I_{\CII} = 7.0 \times 10^{-5}$erg
  $\cmt$s$^{-1}$sr$^{-1}$.  The bottom of the color scale (black)
  shows the high-temperature high density limit from Eq. 2 which
  yields log($N_{C^+}$) $\ge 16.65$ -- in the range of parameter space
  presented here, the limit is not reached.  The column densities are
  for optically thin emission.  Contours are plotted at column
  densities of log(N$_{C^+}$) $= 17$, 17.3, and 17.6.  We have used
  Eq. A1 from \citet{Crawford85} assuming a critical density of
  $3000\cc$.}
\end{flushleft}
\end{figure}

\subsubsection{A more realistic estimate of N$_{C^+}$ }

We can get an idea of how much of an underestimate Eq. (2) is by
looking at Fig.~\ref{ncii}, which shows ionized carbon column density
variations as a function of density and temperature for low optical
depths.  Under these conditions the C$^+$ column density is
proportional to the \CII\ line intensity, so that we can use this
diagnostic diagram to calculate C$^+$ column densities when
information about the temperature and density of the gas is available.
Throughout, the ratio between $N_{C^+}$ and $I_{\CII}$ is at least
twice that of Eq. (2), except in the very high temperature, very high
density regime (the violet region at the upper right).

It is unlikely that the gas (or the dust) in BCLMP~691 has a temperature
T$\gg 91$K and density $n \gg 3000 \cc$ on the (linear) scales 
we are sensitive to here. As noted
before, these scales are almost identical to those \citet{Israel96}
and \citet{Israel11} were sensitive to in the LMC. In
the LMC regions N~159, N~160, and N~11 they estimate $G_0$ to vary
between 25 and 450.  Assuming the flux impinging on a dust cloud is
$I_{star} = 4 \pi I_{dust}$, where the $4 \pi$ converts the
erg$\cmt$s$^{-1}$ sr$^{-1}$ to erg $\cmt$s$^{-1}$, that half of the
stellar flux $I_{star}$ is due to photons with $h\nu < 6$eV
\citep[e.g.][]{Tielens85a}, we can estimate the FUV ($6 < h\nu <
13.6$eV) flux expressed in units of the \citet{Habing68} field to be
$G_0 = 2 \pi I_{dust} / 1.6 \times 10^{-3}$.

Direct integration of the dust emission from 5 to 610~\micron\ (the
edge of the SPIRE 500~\micron\ band) yields a peak flux of 0.013
erg$\cmt$s$^{-1}$ sr$^{-1}$, corresponding to $G_0 = 51$, well within
the range found to apply to the LMC clouds.  Since the core of the
\HII\ region BCLMP~691 is smaller than the $12"$ beam (see H$\alpha$
contours in Fig.~\ref{superpose_spec}, the central $G_0$ is likely to
be somewhat higher but probably lower than in N159 or N160.  
At these temperatures, the dust should emit strongly in the Spitzer
24 and 70\micron\ bands; if the grains are in thermal equilibrium, then a 
temperature of about 90~K is suggested by the flux ratio and the gas 
kinetic temperature should be somewhat higher \citep{Hollenbach99}.
 The strong 5.8\micron\ 
emission shows that some warmer dust is also present. In N159 and N160, 
\citet{Israel96} estimate that the C$^+$ column density is $2
- 4$ times higher than the limit given by Eq. (2).  As $G_0$ rises, we
expect the gas temperature to rise, moving C$^{+}$ column densities
towards the limit expressed by Eq. (2).  Hence, for BCLMP~691, our best
estimate is a column density $N_{C^+} \approx 1.3 \pm 0.4 \times 10^{17}
\cmt$,  3 times the lower limit in Eq. (2).  Below, we confirm this estimate 
using the \citet{Kaufman99} PDR model.

\subsubsection{Hydrogen column density N$_{H}$}

There are no direct measurements of the carbon abundance in
BCLMP~691.  However, \citet{Esteban09} estimate carbon and oxygen 
abundances for NGC~595 and NGC~604 and the C/O ratios they
find are $[C]/[O] = -0.16$ and $-0.20$.  They present models by \citet{Carigi05} 
that suggest C/O ratios of about $[C]/[O] = -0.33$.
Based on the models and these recent and high-quality data, we assume $[C]/[O] = -0.25 \pm 0.08$
yielding a carbon abundance for BCLMP~691 of 12+log[C]/[H] $= 8.17 \pm0.1$, or
a C abundance of $1.48 \times 10^{-4} \pm$25\% after propagating the (independent) 
uncertainties given above for the O abundance and the C/O ratio.
If we assume that half of the carbon is locked 
 in dust grains, this leads us to a gas-phase C abundance of $7.4
\times 10^{-5}$.  Therefore, for $N_{C^+} \approx 1.3 \times 10^{17}
\cmt$, the H column density is about $N_{\rm H} \approx 1.8 \times
10^{21} \cmt$.  If the fraction of carbon locked in dust is 2/3 \citep[as
suggested by Table 2 of ][]{Sofia11} , then the gas-phase C abundance would
be correspondingly lower and the H column density would become $N_{\rm
  H} \approx 2.6 \times 10^{21} \cmt$.

We can compare this column density with that of the HI and the CO.
The \HI\ column density at $12"$ resolution towards the center of BCLMP~691  is 
$N_{\HI} = 1.7 \times 10^{21} \cmt$, of which $1.2 \times 10^{21} \cmt$ is within the
\CII\ velocity range.  Assuming a conversion factor $\ratioo = 4 \times 10^{20} \Xunit$,  twice that of the Galactic disk, the molecular gas column density is $N_H = 2 \NHmol 
\approx 3.2 \times 10^{21} \cmt$.  The \CI\ line has not been observed in BCLMP~691.
At larger scales, the whole \HII\ region emits about 2\% of the H$\alpha$ flux of M~33
and the associated molecular cloud 299, which is about the same size as the \HII\ region 
\citep[see table in ][]{Gratier11}, emits about 0.1\% of the CO emission. 



\section{Analysis of BCLMP~691 \CII\ line profiles}

\subsection{The $^{13}$CII line}

\begin{figure}[!h]
\begin{flushleft}
\includegraphics[angle=0,width=9.5cm]{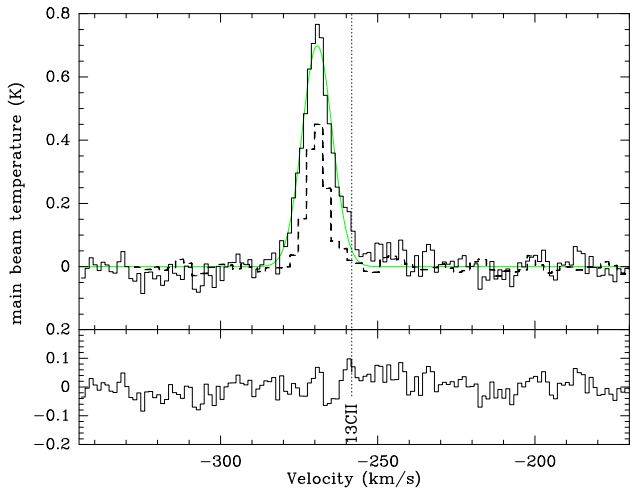}
\caption{\label{13ciib} \CII\ spectrum of the sum of the 5 strongest
  positions of the cross and a gaussian fit (green), showing the
  shoulder at $v\approx -259 \kms$ and the velocity where [$^{13}$\ion{C}{ii}]
  emission would be expected.  The average CO spectrum of these
  positions is shown as a dashed line.  The lower panel shows the
  \CII\ residual (spectrum minus gaussian fit).}
\end{flushleft}
\end{figure}

The three hyperfine components of the [$^{13}$\ion{C}{ii}] line have
frequencies equivalent to offsets of $-66$, $+11$, and $+63\kms$,
respectively, from the [$^{12}$\ion{C}{ii}] line center at
$1900536.9\pm 1.3$ MHz \citep{Cooksy86,Stacey91,Boreiko96}.  The ratio of the sum of the three
hyperfine [$^{13}$\ion{C}{ii}] transitions to the [$^{12}$\ion{C}{ii}]
line intensity, coupled with the relative abundance ratio, provides a
measure of the optical depth of the [$^{12}$\ion{C}{ii}] emission
\citep[e.g.][]{Stacey91}.  We have searched for a possible signature
of [$^{13}$\ion{C}{ii}] emission in the inner regions of BLMP~691
where the $+11\kms$ component may occur as a shoulder present on the
[$^{12}$\ion{C}{ii}] line. After fitting a gaussian to the sum of the
five strongest detections (the 0,0 spectrum and the first spectrum on
each leg of the cross), we find that the
gaussian fit is good on the blue side ($-280\kms$) but leaves an excess on
the red side, at 10--11$\kms$ from the \CII\ line center (see
Fig.~\ref{13ciib}). This transition, $F = 2 \rightarrow 1$, is
expected to be the strongest of the three hyperfine transitions
\citep{Cooksy86}. Although the velocity coverage of the spectra
includes the two other components, they are not detected.
The lower panel of Fig.~\ref{13ciib} shows the residual (spectrum
minus gaussian fit) with the nominal position of the
[$^{13}$\ion{C}{ii}] line.  Using 10 km/s windows shifted by $-66$,
$+11$, and $+63\kms$ with respect to the [$^{12}$\ion{C}{ii}] line
center, we find a nominal summed [$^{13}$\ion{C}{ii}] line intensity of
$0.31\pm 0.23 \K\kms$, i.e. a ratio $97 > I_{12/13} > 16$ with
one sigma uncertainties. 

\begin{figure}[!h]
\begin{flushleft}
\includegraphics[angle=0,width=8.5cm]{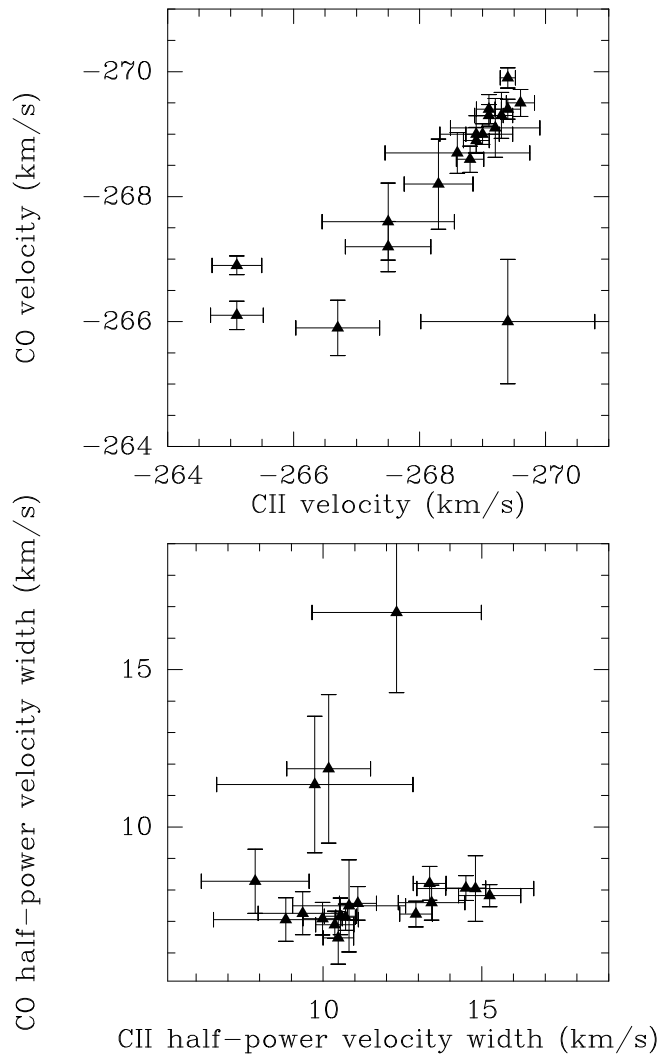}
\caption{\label{vel_disp}  ($top$) Comparison of CO and \CII\ velocities, showing the good agreement.
($bottom$) Comparison of CO and \CII\ line widths, showing that with the exception of some positions with large uncertainties, the CO lines are narrower.  The velocities, line widths, and their uncertainties, have been measured
by gaussian fits to the spectra.}
\end{flushleft}
\end{figure}
\begin{figure*}[!h]
\begin{flushleft}
\includegraphics[angle=0,width=18cm]{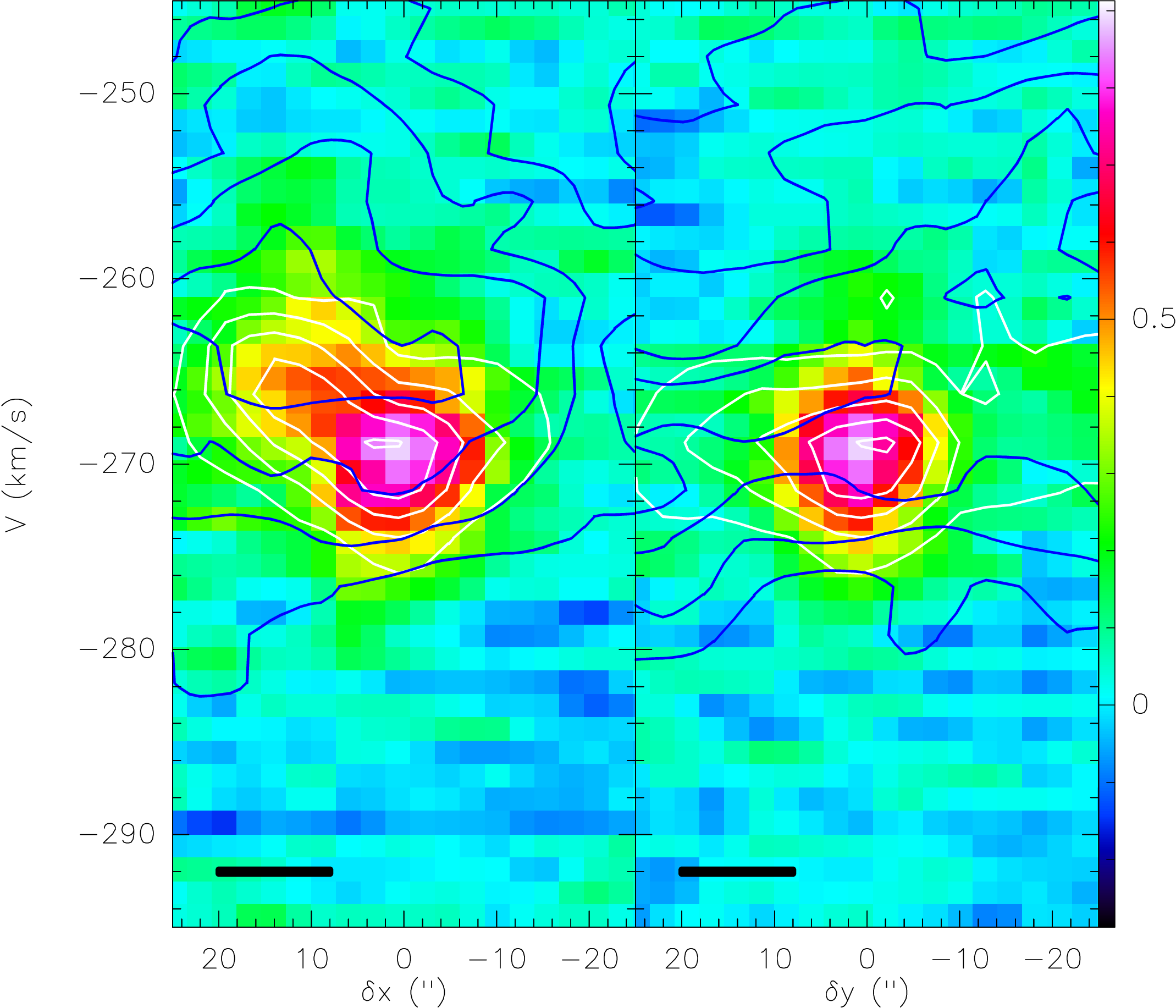}
\caption{\label{pv_plot} Position-velocity plots along the two arms of
  the observed HIFI cross. \emph{Left:} perpendicular to the major
  axis of the galaxy, \emph{Right:} along the major axis of the
  galaxy. For each plot the color bitmap is the [\ion{C}{ii}] in main
  beam temperature (Kelvins), the white contours are the CO(2--1) Tmb
  temperature every 0.1~K starting from 0.1~K, the dark blue contours
  are the \ion{H}{i} temperature every 10~K starting from 10~K.  The
  horizontal bar at the bottom of each panel indicates the 12$"$
  beamsize of the observations.}
\end{flushleft}
\end{figure*}

\subsection{\CII\ line profiles compared to CO and \HI}

As can be seen from both Table 1 and Fig.~\ref{vel_disp}, the CO and
\CII\ radial velocities determined from gaussian fits agree very well,
generally to within half of a CO velocity channel. The exception is
the relatively poor measurement at position (+7.7,+18.5) which is a
($3.5 \sigma$) detection with a very large uncertainty associated with
both the velocity and line width.  In BCLMP~302, Mookerjea et
al. (2011) find a mean velocity difference between CO and \CII\ of
$1.3 \kms$, presumably reflecting physical motion (outflow?) of one of
the gaseous components giving rise to the \CII\ emission.  We do not
find such a velocity difference in BCLMP~691 but that could well be a
question of geometry.  The \CII\ profiles presented here
are similar in width to the HIFI spectrum presented by Mookerjea et
al., as are the CO profiles.

The \CII, \HI, and CO spectra for all positions detected in \CII\ are
presented in Fig.~\ref{spectra}.  While \HI\ is detected everywhere, 
the CO emission peaks in the regions where
\CII\ was detected. The CO and \CII\ profiles are similar in shape
although the \CII\ lines are broader.  \HI, on the other hand, is
systematically shifted to more positive radial velocities ({\it i.e.} 
lower rotation velocities) by about
$5\kms$ and the CO and \CII\ profiles are at the high rotation velocity
edge of the \HI.  Only a few of the \CII\ spectra show emission at
velocities similar to those of the \HI\ (e.g. low rotation 
velocity tail of emission at offsets 9.2,$-3.8$ and 13.9,$-$5.7).

Figure~\ref{pv_plot} presents a comparison of the position-velocity
(PV) plots of \CII, CO, and \HI\ along the two arms of the cross
observed in \CII.  In both directions, the CO temperature and CO velocity
follow those of \CII\ closely but the \HI\ profiles seem
uncorrelated in temperature and spread over a much broader range in
velocities (typically $\sim 20\kms$). 
Figure~\ref{vel_disp} shows the general
agreement in velocities between CO and \CII\ in a different way but it
also suggests (lower panel) that the line widths mostly fall into two
groups.  
The brighter regions (smaller error bars) define a distribution where the CO line
width is essentially constant while the \CII\ line width varies by a
factor two.  At the very least this indicates that the CO-emitting gas
is dynamically cooler, less affected by the processes that cause
large-scale motions in the bright \CII\ emitting region. The weaker
regions, on the other hand, show a pattern where both CO and \CII\ 
appear to vary
linearly with each other. These regions correspond to the more
diffuse ISM. Looking more closely at the bright regions
(i.e. ignoring the regions with larger error bars) we see that there
may be a division into two further groups with \CII\ widths of $\sim
10\kms$ and $\sim 15\kms$ respectively and CO widths of $\sim 7\kms$
and $\sim 8\kms$ respectively.  The broad and strong \CII\ lines are found
along the H$\alpha$-bright extension to the SE.  

\section{Synthesis and Conclusions}

In order to better locate BCLMP~691 on the diagnostic diagram in
Fig.~\ref{ncii}, we need an estimate of the temperature, density, or a
combination.  We can place the \CII/FIR ratio of $\sim 0.011$ derived
in Sect. 3.2 on Fig. 11 in \citet{Bakes94} and find that the product
$G_0 \sqrt{T} / n_e \approx 20000$.  For a neutral region, the
electrons will come from the singly ionized carbon so $n_e \approx n_H
\, 7.4 \times 10^{-5}$ (see Sect. 3.3.3).  Noting $G_2 = G_0 / 100$, $T=100 T_2$, and
$n_H = 1000 n_3$, we obtain $G_2 \sqrt{T_2} / n_3 \approx 1.5$.
In Sect. 3.1.2 we estimated $G_0 \sim 50$ but pointed out that this
could be an underestimate by a factor  2.
\citet{Kaufman99} estimate gas temperatures varying from 300~K to
100~K for $G_0 = 100$ as densities climb from 100 to 10000$\cc$, with
$n_H = 1000$ corresponding to $T \approx 200$~K.

The general similarity between BCLMP~691 and BCLMP~302 holds also for
the \CII /\OI\ ratio, which is \CII /\OI\  $\ga 1$ in both regions
\citep[see Table 7 in ] [and Nikola et al. in prep.]{Mookerjea11}.
Looking at e.g. Fig. 4 of \citet{Kaufman99}, we see that for a
constant \CII /\OI\ ratio, $n_H$ decreases when $T$ increases.  The
"solution" $G_0 \sim 100, n_H \sim 1000$ seems to be a good fit and the
value of $T \sim 200$K fits both with the \citet{Kaufman99} PDR model and
with the \CII/FIR ratio via the calculations in \citet{Bakes94}.
Looking back at Fig.~\ref{ncii}, the \CII\ column density is about
$N_{\CII} \sim 1.3 \times 10^{17} \cmt$, in good agreement with
Sect. 3.3.2.  It may be worth noting that some "extra" H$_2$ was found by
\citet{Braine10a} through correlating the Herschel/SPIRE cool dust
measurements with CO and \HI\ observations.  Their interpretation was
indeed that this column density could likely be attributed to regions,
cloud envelopes, where the CO was photo-dissociated but the gas remained
predominantly molecular.  The $1.3 \pm \times 10^{17} \cmt$ C$^+$ column we observe
comes from this layer of \HI\ and H$_2$ exterior to the region where most carbon
is in CO.  As the C$^+$ layer contains more H than is observed in \HI , particularly
when considering only the \HI\ within the velocity range detected in \CII , some 
molecular gas not detected via CO must be present.



The correspondence along the cross between the \CII\ and CO
intensities and velocities is quite good (see e.g. Fig.~\ref{spectra})
in BCLMP~691 but this is not the case for all regions.  We suspect
that this may be a question of geometry and that in BCLMP~691 we are
probably seeing the \HII\ region with the molecular cloud right behind
it such that the surface of the cloud is in the same direction as the
bulk of the cloud (where carbon is in CO).  At the scale of this
\HII\ region, the \CII\ emission is clearly coming from the PDR, plus
perhaps the ionized skin.  BCLMP~691 appears to be part of molecular
cloud 299 in the \citet{Gratier11} catalog.  The strong emission in
all wavebands tracing star formation, including the FUV, suggests that
indeed the \HII\ region is on the near side of the GMC.  While we use
the term "cloud", molecular clouds have substructure so the exposed
material may have a rather intricate perimeter. The CO peak is
actually just south-east of the H$\alpha$ peak (the \HII\ region) but
the FUV, 8, and 24~\micron\ emission all peak on the \HII\ region,
implying that the warm part of the molecular cloud is behind the
\HII\ region.  Since the \CII\ emission is coming from the near side
of the molecular cloud, it is not surprising that it shares the same
dynamics (velocities, Fig.~\ref{vel_disp}).  Perhaps due to the
geometry, these observations of BCLMP~691 show clearly that the
\CII\ emission is coming from the molecular cloud -- \HII\ region
interface, as in models of photo-dissociation regions
\citep{Tielens85a,Kaufman99}.


\begin{acknowledgements}
The authors would like to thank Laura Magrini for useful discussions on abundances in M33. 

HIFI has been designed and built by a consortium of
institutes and university departments from across Europe, Canada, and the
United States under the leadership of SRON Netherlands Institute for Space
Research, Groningen, The Netherlands, and with major contributions from
Germany, France, and the US. Consortium members are: Canada: CSA,
U. Waterloo; France: CESR, LAB, LERMA, IRAM; Germany: KOSMA,
MPIfR, MPS; Ireland, NUI Maynooth; Italy: ASI, IFSI-INAF, Osservatorio
Astrofisico di Arcetri-INAF; Netherlands: SRON, TUD; Poland: CAMK, CBK;
Spain: Observatorio Astron\'omico Nacional (IGN), Centro de Astrobiolog\'{i}a
(CSIC-INTA). Sweden: Chalmers University of Technology Ð MC2, RSS \&
GARD; Onsala Space Observatory; Swedish National Space Board, Stockholm
University Ð Stockholm Observatory; Switzerland: ETH Zurich, FHNW; USA:
Caltech, JPL, NHSC. HIPE is a joint development by the Herschel Science
Ground Segment Consortium, consisting of ESA, the NASA Herschel Science
Center, and the HIFI, PACS and SPIRE consortia. We also thank the French
Space Agency CNES for financial support.
\end{acknowledgements}

\bibliographystyle{aa}
\bibliography{jb}

\begin{thebibliography}{44}
\expandafter\ifx\csname natexlab\endcsname\relax\def\natexlab#1{#1}\fi

\bibitem[{{Bakes} \& {Tielens}(1994)}]{Bakes94}
{Bakes}, E.~L.~O. \& {Tielens}, A.~G.~G.~M. 1994, ApJ, 427, 822

\bibitem[{{Boreiko} \& {Betz}(1991)}]{Boreiko91}
{Boreiko}, R.~T. \& {Betz}, A.~L. 1991, ApJL, 380, L27

\bibitem[{{Boreiko} \& {Betz}(1996)}]{Boreiko96}
{Boreiko}, R.~T. \& {Betz}, A.~L. 1996, ApJL, 467, L113+

\bibitem[{{Boulesteix} {et~al.}(1974){Boulesteix}, {Courtes}, {Laval},
  {Monnet}, \& {Petit}}]{Boulesteix74}
{Boulesteix}, J., {Courtes}, G., {Laval}, A., {Monnet}, G., \& {Petit}, H.
  1974, A\&A, 37, 33

\bibitem[{{Braine} {et~al.}(2010){Braine}, {Gratier}, {Kramer}, {Schuster},
  {Tabatabaei}, \& {Gardan}}]{Braine10a}
{Braine}, J., {Gratier}, P., {Kramer}, C., {et~al.} 2010, A\&A, 520, A107+

\bibitem[{{Braine} \& {Hughes}(1999)}]{Braine_n4414c}
{Braine}, J. \& {Hughes}, D.~H. 1999, A\&A, 344, 779

\bibitem[{{Carigi} {et~al.}(2005){Carigi}, {Peimbert}, {Esteban}, \&
  {Garc{\'{\i}}a-Rojas}}]{Carigi05}
{Carigi}, L., {Peimbert}, M., {Esteban}, C., \& {Garc{\'{\i}}a-Rojas}, J. 2005,
  ApJ, 623, 213

\bibitem[{{Cooksy} {et~al.}(1986){Cooksy}, {Blake}, \& {Saykally}}]{Cooksy86}
{Cooksy}, A.~L., {Blake}, G.~A., \& {Saykally}, R.~J. 1986, ApJL, 305, L89

\bibitem[{{Crawford} {et~al.}(1985){Crawford}, {Genzel}, {Townes}, \&
  {Watson}}]{Crawford85}
{Crawford}, M.~K., {Genzel}, R., {Townes}, C.~H., \& {Watson}, D.~M. 1985, ApJ,
  291, 755

\bibitem[{{Dale} \& {Helou}(2002)}]{Dale02}
{Dale}, D.~A. \& {Helou}, G. 2002, ApJ, 576, 159

\bibitem[{{de Graauw} {et~al.}(2010){de Graauw}, {Helmich}, {Phillips},
  {Stutzki}, {Caux}, {Whyborn}, {Dieleman}, {Roelfsema}, {Aarts}, {Assendorp},
  {Bachiller}, {Baechtold}, {Barcia}, {Beintema}, {Belitsky}, {Benz}, {Bieber},
  {Boogert}, {Borys}, {Bumble}, {Ca{\"i}s}, {Caris}, {Cerulli-Irelli},
  {Chattopadhyay}, {Cherednichenko}, {Ciechanowicz}, {Coeur-Joly}, {Comito},
  {Cros}, {de Jonge}, {de Lange}, {Delforges}, {Delorme}, {den Boggende},
  {Desbat}, {Diez-Gonz{\'a}lez}, {di Giorgio}, {Dubbeldam}, {Edwards},
  {Eggens}, {Erickson}, {Evers}, {Fich}, {Finn}, {Franke}, {Gaier}, {Gal},
  {Gao}, {Gallego}, {Gauffre}, {Gill}, {Glenz}, {Golstein}, {Goulooze},
  {Gunsing}, {G{\"u}sten}, {Hartogh}, {Hatch}, {Higgins}, {Honingh}, {Huisman},
  {Jackson}, {Jacobs}, {Jacobs}, {Jarchow}, {Javadi}, {Jellema}, {Justen},
  {Karpov}, {Kasemann}, {Kawamura}, {Keizer}, {Kester}, {Klapwijk}, {Klein},
  {Kollberg}, {Kooi}, {Kooiman}, {Kopf}, {Krause}, {Krieg}, {Kramer},
  {Kruizenga}, {Kuhn}, {Laauwen}, {Lai}, {Larsson}, {Leduc}, {Leinz}, {Lin},
  {Liseau}, {Liu}, {Loose}, {L{\'o}pez-Fernandez}, {Lord}, {Luinge}, {Marston},
  {Mart{\'{\i}}n-Pintado}, {Maestrini}, {Maiwald}, {McCoey}, {Mehdi}, {Megej},
  {Melchior}, {Meinsma}, {Merkel}, {Michalska}, {Monstein}, {Moratschke},
  {Morris}, {Muller}, {Murphy}, {Naber}, {Natale}, {Nowosielski}, {Nuzzolo},
  {Olberg}, {Olbrich}, {Orfei}, {Orleanski}, {Ossenkopf}, {Peacock}, {Pearson},
  {Peron}, {Phillip-May}, {Piazzo}, {Planesas}, {Rataj}, {Ravera}, {Risacher},
  {Salez}, {Samoska}, {Saraceno}, {Schieder}, {Schlecht}, {Schl{\"o}der},
  {Schm{\"u}lling}, {Schultz}, {Schuster}, {Siebertz}, {Smit}, {Szczerba},
  {Shipman}, {Steinmetz}, {Stern}, {Stokroos}, {Teipen}, {Teyssier}, {Tils},
  {Trappe}, {van Baaren}, {van Leeuwen}, {van de Stadt}, {Visser}, {Wildeman},
  {Wafelbakker}, {Ward}, {Wesselius}, {Wild}, {Wulff}, {Wunsch}, {Tielens},
  {Zaal}, {Zirath}, {Zmuidzinas}, \& {Zwart}}]{deGraauw10}
{de Graauw}, T., {Helmich}, F.~P., {Phillips}, T.~G., {et~al.} 2010, A\&A, 518,
  L6+

\bibitem[{{Esteban} {et~al.}(2009){Esteban}, {Bresolin}, {Peimbert},
  {Garc{\'{\i}}a-Rojas}, {Peimbert}, \& {Mesa-Delgado}}]{Esteban09}
{Esteban}, C., {Bresolin}, F., {Peimbert}, M., {et~al.} 2009, ApJ, 700, 654

\bibitem[{{Gratier} {et~al.}(2010){Gratier}, {Braine}, {Rodriguez-Fernandez},
  {Schuster}, {Kramer}, {Xilouris}, {Tabatabaei}, {Henkel}, {Corbelli},
  {Israel}, {van der Werf}, {Calzetti}, {Garcia-Burillo}, {Sievers}, {Combes},
  {Wiklind}, {Brouillet}, {Herpin}, {Bontemps}, {Aalto}, {Koribalski}, {van der
  Tak}, {Wiedner}, {Roellig}, \& {Mookerjea}}]{Gratier10}
{Gratier}, P., {Braine}, J., {Rodriguez-Fernandez}, N.~J., {et~al.} 2010, A\&A,
  522, A3

\bibitem[{{Gratier} {et~al.}(2011){Gratier}, {Braine}, \&
  {Rodriguez-Fernandez}}]{Gratier11}
{Gratier}, P., {Braine}, J., \& {Rodriguez-Fernandez}, N.~J. e.~a. 2011, A\&A,
  00, 00

\bibitem[{{Greenawalt}(1998)}]{Greenawalt98}
{Greenawalt}, B.~E. 1998, PhD thesis, New Mexico State University

\bibitem[{{Habing}(1968)}]{Habing68}
{Habing}, H.~J. 1968, \bain, 19, 421

\bibitem[{{Heiles}(1994)}]{Heiles94}
{Heiles}, C. 1994, ApJ, 436, 720

\bibitem[{{Hollenbach} \& {Tielens}(1999)}]{Hollenbach99}
{Hollenbach}, D.~J. \& {Tielens}, A.~G.~G.~M. 1999, Reviews of Modern Physics,
  71, 173

\bibitem[{{Hoopes} \& {Walterbos}(2000)}]{Hoopes00}
{Hoopes}, C.~G. \& {Walterbos}, R.~A.~M. 2000, ApJ, 541, 597

\bibitem[{{Israel} \& {Maloney}(2011)}]{Israel11}
{Israel}, F.~P. \& {Maloney}, P.~R. 2011, A\&A, 531, A19

\bibitem[{{Israel} {et~al.}(1996){Israel}, {Maloney}, {Geis}, {Herrmann},
  {Madden}, {Poglitsch}, \& {Stacey}}]{Israel96}
{Israel}, F.~P., {Maloney}, P.~R., {Geis}, N., {et~al.} 1996, ApJ, 465, 738

\bibitem[{{Israel} \& {van der Kruit}(1974)}]{Israel74}
{Israel}, F.~P. \& {van der Kruit}, P.~C. 1974, A\&A, 32, 363

\bibitem[{{Kaufman} {et~al.}(1999){Kaufman}, {Wolfire}, {Hollenbach}, \&
  {Luhman}}]{Kaufman99}
{Kaufman}, M.~J., {Wolfire}, M.~G., {Hollenbach}, D.~J., \& {Luhman}, M.~L.
  1999, ApJ, 527, 795

\bibitem[{{Kramer} {et~al.}(2010){Kramer}, {Buchbender}, {Xilouris}, {Boquien},
  {Braine}, {Calzetti}, {Lord}, {Mookerjea}, {Quintana-Lacaci}, {Rela{\~n}o},
  {Stacey}, {Tabatabaei}, {Verley}, {Aalto}, {Akras}, {Albrecht}, {Anderl},
  {Beck}, {Bertoldi}, {Combes}, {Dumke}, {Garcia-Burillo}, {Gonzalez},
  {Gratier}, {G{\"u}sten}, {Henkel}, {Israel}, {Koribalski}, {Lundgren},
  {Martin-Pintado}, {R{\"o}llig}, {Rosolowsky}, {Schuster}, {Sheth}, {Sievers},
  {Stutzki}, {Tilanus}, {van der Tak}, {van der Werf}, \& {Wiedner}}]{Kramer10}
{Kramer}, C., {Buchbender}, C., {Xilouris}, E.~M., {et~al.} 2010, A\&A, 518,
  L67

\bibitem[{{Langer} {et~al.}(2010){Langer}, {Velusamy}, {Pineda}, {Goldsmith},
  {Li}, \& {Yorke}}]{Langer10}
{Langer}, W.~D., {Velusamy}, T., {Pineda}, J.~L., {et~al.} 2010, A\&A, 521,
  L17+

\bibitem[{{Loeb}(1993)}]{Loeb93}
{Loeb}, A. 1993, ApJL, 404, L37

\bibitem[{{Magrini} {et~al.}(2010){Magrini}, {Stanghellini}, {Corbelli},
  {Galli}, \& {Villaver}}]{Magrini10}
{Magrini}, L., {Stanghellini}, L., {Corbelli}, E., {Galli}, D., \& {Villaver},
  E. 2010, A\&A, 512, 63

\bibitem[{{Maiolino} {et~al.}(2005){Maiolino}, {Cox}, {Caselli}, {Beelen},
  {Bertoldi}, {Carilli}, {Kaufman}, {Menten}, {Nagao}, {Omont}, {Wei{\ss}},
  {Walmsley}, \& {Walter}}]{Maiolino05}
{Maiolino}, R., {Cox}, P., {Caselli}, P., {et~al.} 2005, A\&A, 440, L51

\bibitem[{{Malhotra} {et~al.}(2001){Malhotra}, {Kaufman}, {Hollenbach},
  {Helou}, {Rubin}, {Brauher}, {Dale}, {Lu}, {Lord}, {Stacey}, {Contursi},
  {Hunter}, \& {Dinerstein}}]{Malhotra01}
{Malhotra}, S., {Kaufman}, M.~J., {Hollenbach}, D., {et~al.} 2001, ApJ, 561,
  766

\bibitem[{{Mookerjea} {et~al.}(2011){Mookerjea}, {Kramer}, {Buchbender},
  {Boquien}, {Verley}, {Rela{\~n}o}, {Quintana-Lacaci}, {Aalto}, {Braine},
  {Calzetti}, {Combes}, {Garcia-Burillo}, {Gratier}, {Henkel}, {Israel},
  {Lord}, {Nikola}, {R{\"o}llig}, {Stacey}, {Tabatabaei}, {van der Tak}, \&
  {van der Werf}}]{Mookerjea11}
{Mookerjea}, B., {Kramer}, C., {Buchbender}, C., {et~al.} 2011, A\&A, 532, A152

\bibitem[{{Pak} {et~al.}(1998){Pak}, {Jaffe}, {van Dishoeck}, {Johansson}, \&
  {Booth}}]{Pak98}
{Pak}, S., {Jaffe}, D.~T., {van Dishoeck}, E.~F., {Johansson}, L.~E.~B., \&
  {Booth}, R.~S. 1998, ApJ, 498, 735

\bibitem[{{Papadopoulos} {et~al.}(2002){Papadopoulos}, {Thi}, \&
  {Viti}}]{Papadopoulos02}
{Papadopoulos}, P.~P., {Thi}, W.-F., \& {Viti}, S. 2002, ApJ, 579, 270

\bibitem[{{Pfenniger} {et~al.}(1994){Pfenniger}, {Combes}, \&
  {Martinet}}]{Pfenniger94a}
{Pfenniger}, D., {Combes}, F., \& {Martinet}, L. 1994, A\&A, 285, 79

\bibitem[{{Pilbratt}(2010)}]{Pilbratt10}
{Pilbratt}, G. e.~a. 2010, A\&A, this, volume

\bibitem[{{Poglitsch} {et~al.}(1995){Poglitsch}, {Krabbe}, {Madden}, {Nikola},
  {Geis}, {Johansson}, {Stacey}, \& {Sternberg}}]{Poglitsch95}
{Poglitsch}, A., {Krabbe}, A., {Madden}, S.~C., {et~al.} 1995, ApJ, 454, 293

\bibitem[{{Rodr{\'{\i}}guez-Fern{\'a}ndez}
  {et~al.}(2006){Rodr{\'{\i}}guez-Fern{\'a}ndez}, {Braine}, {Brouillet}, \&
  {Combes}}]{Rodriguez06}
{Rodr{\'{\i}}guez-Fern{\'a}ndez}, N.~J., {Braine}, J., {Brouillet}, N., \&
  {Combes}, F. 2006, A\&A, 453, 77

\bibitem[{{Roelfsema} {et~al.}(2012){Roelfsema}, {Helmich}, {Teyssier},
  {Ossenkopf}, {Morris}, {Olberg}, {Shipman}, {Risacher}, {Akyilmaz},
  {Assendorp}, {Avruch}, {Beintema}, {Biver}, {Boogert}, {Borys}, {Braine},
  {Caris}, {Caux}, {Cernicharo}, {Coeur-Joly}, {Comito}, {de Lange},
  {Delforge}, {Dieleman}, {Dubbeldam}, {de Graauw}, {Edwards}, {Fich},
  {Flederus}, {Gal}, {di Giorgio}, {Herpin}, {Higgins}, {Hoac}, {Huisman},
  {Jarchow}, {Jellema}, {de Jonge}, {Kester}, {Klein}, {Kooi}, {Kramer},
  {Laauwen}, {Larsson}, {Leinz}, {Lord}, {Lorenzani}, {Luinge}, {Marston},
  {Mart{\'{\i}}n-Pintado}, {McCoey}, {Melchior}, {Michalska}, {Moreno},
  {M{\"u}ller}, {Nowosielski}, {Okada}, {Orlea{\'n}ski}, {Phillips}, {Pearson},
  {Rabois}, {Ravera}, {Rector}, {Rengel}, {Sagawa}, {Salomons},
  {S{\'a}nchez-Su{\'a}rez}, {Schieder}, {Schl{\"o}der}, {Schm{\"u}lling},
  {Soldati}, {Stutzki}, {Thomas}, {Tielens}, {Vastel}, {Wildeman}, {Xie},
  {Xilouris}, {Wafelbakker}, {Whyborn}, {Zaal}, {Bell}, {Bjerkeli}, {De Beck},
  {Cavali{\'e}}, {Crockett}, {Hily-Blant}, {Kama}, {Kaminski}, {Lefl{\'o}ch},
  {Lombaert}, {de Luca}, {Makai}, {Marseille}, {Nagy}, {Pacheco}, {van der
  Wiel}, {Wang}, \& {Y{\i}ld{\i}z}}]{Roelfsema12}
{Roelfsema}, P.~R., {Helmich}, F.~P., {Teyssier}, D., {et~al.} 2012, A\&A, 537,
  A17

\bibitem[{{Rubin} {et~al.}(2008){Rubin}, {Simpson}, {Colgan}, {Dufour},
  {Brunner}, {McNabb}, {Pauldrach}, {Erickson}, {Haas}, \& {Citron}}]{Rubin08}
{Rubin}, R.~H., {Simpson}, J.~P., {Colgan}, S.~W.~J., {et~al.} 2008, MNRAS,
  387, 45

\bibitem[{{Sofia} {et~al.}(2011){Sofia}, {Parvathi}, {Babu}, \&
  {Murthy}}]{Sofia11}
{Sofia}, U.~J., {Parvathi}, V.~S., {Babu}, B.~R.~S., \& {Murthy}, J. 2011, AJ,
  141, 22

\bibitem[{{Stacey} {et~al.}(1991){Stacey}, {Geis}, {Genzel}, {Lugten},
  {Poglitsch}, {Sternberg}, \& {Townes}}]{Stacey91}
{Stacey}, G.~J., {Geis}, N., {Genzel}, R., {et~al.} 1991, ApJ, 373, 423

\bibitem[{{Tabatabaei} {et~al.}(2007){Tabatabaei}, {Beck}, {Krause},
  {Berkhuijsen}, {Gehrz}, {Gordon}, {Hinz}, {Humphreys}, {McQuinn}, {Polomski},
  {Rieke}, \& {Woodward}}]{Tabatabaei07a}
{Tabatabaei}, F.~S., {Beck}, R., {Krause}, M., {et~al.} 2007, A\&A, 466, 509

\bibitem[{{Tielens} \& {Hollenbach}(1985)}]{Tielens85a}
{Tielens}, A.~G.~G.~M. \& {Hollenbach}, D. 1985, ApJ, 291, 722

\bibitem[{{Verley} {et~al.}(2007){Verley}, {Hunt}, {Corbelli}, \&
  {Giovanardi}}]{Verley07}
{Verley}, S., {Hunt}, L.~K., {Corbelli}, E., \& {Giovanardi}, C. 2007, A\&A,
  476, 1161

\bibitem[{{Wolfire} {et~al.}(2010){Wolfire}, {Hollenbach}, \&
  {McKee}}]{Wolfire10}
{Wolfire}, M.~G., {Hollenbach}, D., \& {McKee}, C.~F. 2010, ApJ, 716, 1191

\end{thebibliography}

\end{document}